\documentclass[conference]{IEEEtran}
\IEEEoverridecommandlockouts
\usepackage{cite}
\usepackage{amsmath,amssymb,amsfonts}
\usepackage{hyperref}
\usepackage[dvipsnames]{xcolor}

\usepackage[colorinlistoftodos,prependcaption,textsize=tiny]{todonotes}
\usepackage{microtype}
\usepackage[utf8]{inputenc}
\usepackage{graphicx}
\usepackage{textcomp}
\usepackage{xcolor}
\usepackage{subfigure}

\usepackage[noend]{algpseudocode}
\usepackage{tabularx}
\usepackage{booktabs}
\usepackage{paralist}
\usepackage{etoolbox}
\usepackage{enumitem}

\usepackage{rotating}
\usepackage{booktabs}
\usepackage{multirow}
\usepackage{verbatim }
\usepackage{booktabs,graphicx}
\usepackage[ruled,linesnumbered]{algorithm2e}

\newtheorem{theorem}{Theorem}[section]
\newtheorem{proposition}{Proposition}
\newtheorem{lemma}[theorem]{Lemma}
\newtheorem{definition}{Definition}[section]

\usepackage[colorinlistoftodos,prependcaption,textsize=tiny]{todonotes}

\usepackage{etoolbox}
\usepackage{enumitem}
\usepackage{array, boldline, rotating}

\hypersetup{
	pdftoolbar=true,        
	pdfmenubar=true,        
	pdffitwindow=false,     
	pdfstartview={FitH},    
	pdftitle={Libra: High-Utility Anonymization of Event Logs for Process Mining via Subsampling},    
	pdfauthor={},     
	pdfsubject={},   
	pdfcreator={},   
	pdfproducer={}, 
	pdfkeywords={}, 
	pdfnewwindow=true,      
	colorlinks=true,       
	linkcolor=Brown,          
	citecolor=OliveGreen,        
	filecolor=magenta,      
	urlcolor=NavyBlue           
}

\def\BibTeX{{\rm B\kern-.05em{\sc i\kern-.025em b}\kern-.08em
    T\kern-.1667em\lower.7ex\hbox{E}\kern-.125emX}}
    
\begin{document}

\title{Libra: High-Utility Anonymization of Event Logs for Process Mining via Subsampling\thanks{Work funded by the European Research Council (PIX project)}}

\author{\IEEEauthorblockN{Gamal Elkoumy}
\IEEEauthorblockA{
\textit{University of Tartu}, 
Tartu, Estonia \\
gamal.elkoumy@ut.ee}
\and
\IEEEauthorblockN{Marlon Dumas}
\IEEEauthorblockA{
\textit{University of Tartu}, 
Tartu, Estonia \\
marlon.dumas@ut.ee}

}

\maketitle

\begin{abstract}
Process mining techniques enable analysts to identify and assess process improvement opportunities based on event logs.
A common roadblock to process mining is that event logs may contain private information that cannot be used for analysis without consent.
An approach to overcome this roadblock is to anonymize the event log so that no individual represented in the original log can be singled out based on the anonymized one.
Differential privacy is an anonymization approach that provides this guarantee.
A differentially private event log anonymization technique seeks to produce an anonymized log that is as similar as possible to the original one (high utility) while providing a required privacy guarantee.
Existing event log anonymization techniques operate by injecting noise into the traces in the log (e.g., duplicating, perturbing, or filtering out some traces).
Recent work on differential privacy has shown that a better privacy-utility tradeoff can be achieved by applying subsampling prior to noise injection. In other words, subsampling amplifies privacy.
This paper proposes an event log anonymization approach called Libra that exploits this observation. Libra extracts multiple samples of traces from a log, independently injects noise, retains statistically relevant traces from each sample, and composes the samples to produce a differentially private log. 
An empirical evaluation shows that the proposed approach leads to a considerably higher utility for equivalent privacy guarantees relative to existing baselines.
\end{abstract}


\vspace*{-2mm}
\section{Introduction}
\label{sec:intro}

Process Mining is a family of techniques to analyze event logs generated by enterprise information systems to help organizations identify opportunities to enhance their operational efficiency, compliance, and quality of service~\cite{dumas2013fundamentals}. The primary input of a process mining technique is an event log. 
Each event represents the execution of an instance of an activity. An event contains a reference to a process instance (case identifier), a reference to an activity label, and at least one timestamp. An event may include other attributes, such as the resource (e.g., worker) who performed the activity. Table~\ref{tbl:event_log} presents an excerpt of a healthcare process event log.

Event logs may contain private information. For example, an adversary (or a data analyst) may use the sequence of activities performed in a case to single out patients. For example, patient 5 in Table~\ref{tbl:event_log} is the only case with the sequence $<$Registration, Antibiotics, Release$>$; given that Alice is the only patient who got antibiotics treatment in the log, the adversary can link case 5 to Alice. Also, the correlation between activities and their execution timestamp could be used to single out patients. For example, patient 1 in Table~\ref{tbl:event_log} is the only patient who registered in the morning on the 8th of April; since Bob is the only patient who was at the registration disk in the morning on that day, the adversary can link case 1 to Bob.
Thus, given a released log, an adversary can perform a re-identification attack and single out an individual
~\cite{elkoumy2021privacy}. 

\begin{table}[hbtp]
\vspace*{-4mm}
\centering
\caption{Example of an event log}
\vspace*{-3mm}
\scriptsize
	\begin{tabular}[t]{|c|c|c|c|}
	
\hline

Case ID	&	Activity	&	Timestamp	&	Other Attributes	\\ \hline
\multirow{3}{*}{1}	&	Registration	&	4/8/2021 10:20	&	….....	\\
	&	Triage	&	4/8/2021 10:50	&	….....	\\
	&	Admission	&	4/8/2021 16:15	&	….....	\\ \hline
\multirow{4}{*}{2}	&	Registration	&	4/8/2021 12:37	&	….....	\\ 
	&	Admission	&	4/8/2021 14:37	&	….....	\\
	&	Surgery	&	4/8/2021 15:07	&	….....	\\
	&	Release	&	4/8/2021 20:31	&	….....	\\\hline
\multirow{3}{*}{3}	&	Registration	&	4/9/2021 13:30	&	….....	\\
	&	Triage	&	4/9/2021 13:55	&	….....	\\
	&	Admission	&	4/9/2021 20:55	&	….....	\\\hline
\multirow{4}{*}{4}	&	Registration	&	4/9/2021 15:00	&	….....	\\
	&	Admission	&	4/9/2021 17:00	&	….....	\\
	&	Surgery	&	4/9/2021 17:40	&	….....	\\
	&	Release	&	4/9/2021 23:05	&	….....	\\ \hline
\multirow{3}{*}{5}	&	Registration	&	4/9/2021 17:25	&	….....	\\
	&	Antibiotics	&	4/9/2021 17:55	&	….....	\\
	&	Release	&	4/10/2021 23:55	&	….....	\\ \hline

	\end{tabular}

\label{tbl:event_log}
\vspace*{-3mm}
\end{table}

Data regulations, such as GDPR\footnote{\url{http://data.europa.eu/eli/reg/2016/679/oj}}, require organizations to put in place mechanisms to protect information about individuals when processing a dataset. One way to address this requirement is by using \textit{Differential Privacy} (DP). DP stands out among other Privacy-Enhancing Techniques (PETs) due to its composability and privacy guarantees~\cite{dwork2014algorithmic}. In a nutshell, DP is a guarantee to each individual that the ability for the analyst to infer private information about them will likely be the same whether their record(s)  is/are part of the dataset or not. In other words, DP prevents singling out an individual after disclosing the data~\cite{cohen2020towards}.
DP works by injecting noise into the data. This noise is quantified by a \emph{privacy budget} parameter called $\epsilon$, which captures the extent to which the presence or absence of the record(s) associated to an individual person in the original data, affect the disclosed (anonymized) data. The smaller the $\epsilon$ value, the stronger the privacy guarantee and the larger the injected noise.
Thus, by controlling the privacy budget, we can ensure that the resulting anonymized dataset provides privacy guarantees, while still being useful for analysis. For example, in a healthcare event log, we can control the $\epsilon$ privacy budget to ensure that we can draw conclusions about bottlenecks and improvement opportunities from the anonymized log, while providing privacy guarantees to the patients whose clinical trajectories are represented by the traces in the log.

In this setting, this paper addresses the following problem:  \noindent\textit{Given an event log L, wherein each trace contains private information about an individual (e.g.\ a customer), and given a privacy budget $\epsilon$, generate an anonymized event log L$'$ that provides an $\epsilon$-differential privacy guarantee to each individual represented in the log}.





Existing approaches for event log anonymization based on differential privacy~\cite{mannhardt2019privacy,fahrenkrog2021sacofa,fahrenkrog2020pripel,Elkoumy21MineMe} inject noise to anonymize the frequency distribution of distinct sequences of activities in a log (a.k.a.\ the trace variants) and the event timestamps. This noise injection may introduce behavior not observed in the original log. The extent to which a DP-anonymization technique distorts the data is called the \emph{utility loss}. The holy grail of anonymization techniques in general, and DP-anonymization techniques in particular, is to achieve a low level of re-identification risk with low utility loss.


Recent work in the field of DP has shown that the privacy guarantees of a differentially private mechanism can be amplified by applying it to a small random subsample of records~\cite{ballePrivacyAmplification}. 
This property is known as \textit{privacy amplification}. The underpinning idea is that there is less utility loss overall since we inject less noise into the subsampled records.
In this paper, we hypothesize that a DP approach based on subsampling can achieve lower utility loss for a given level of privacy guarantee relative to existing DP-anonymization techniques for event logs, which are based purely on noise injection.


The contribution of the paper is a DP-anonymization approach for event logs, namely Libra.
Libra starts by filtering out trace variants that, if disclosed, would lead to privacy breaches.
It then extracts multiple Poisson subsamples and applies a DP mechanism to anonymize each subsample.
The resulting differentially private subsamples are combined to construct an anonymized log.
Using the differential privacy composition theorem~\cite{dworkComposition} and the privacy amplification results associated with \textit{Renyi Differential Privacy} (RDP)~\cite{ZhuW19Poission}, we estimate the amplified $\epsilon'$ privacy guarantee provided by the anonymized log. The paper reports on an empirical evaluation to assess the utility loss resulting from the anonymization procedure w.r.t the task of discovering the process map (Directly-Follows Graph) of an event log. The evaluation compares Libra against state-of-the-art approaches using eight real-life logs.



The paper is structured as follows. Sect.~\ref{sec:background} introduces the background and related work. 
Sect.~\ref{sec:approach} presents the anonymization approach. Sect.~\ref{sec:experiment} presents an empirical evaluation. Finally, Sect.~\ref{sec:conclusion} concludes and discusses future work.

\section{Background and Related Work}
This section introduces differential privacy (DP) and formalizes its definition. We then overview the related work of privacy-preserving process mining (PPPM).
\label{sec:background}

\subsection{Differential Privacy (DP)}
\label{sec:diff_priv}

A dataset $D$ consists of data points (entire cases in the event logs) that are drawn from a universe $U$. Two data sets $D$, and $D'$ are called neighboring (adjacent) data sets if they differ in one data point at most. The two data sets $D, D'$ can differ in: (1) the existence of a data point, which is called unbounded differential privacy~\cite{dwork2006}; (2) the value of a data point, which is called bounded differential privacy~\cite{dwork2006calibrating}.

\begin{definition}[Unbounded $(\epsilon, \delta)$-Differential Privacy)~\cite{dwork2006}] \label{def:dp}
A mechanism M is $(\epsilon, \delta)$ differentially private if every neighboring pair of datasets $D, D' \in U$ differing at most on one data point, and all $S \subseteq Range (M)$, the following inequality holds:
$Pr[M(D) \in S] \leq e^\epsilon \cdot Pr[M(D') \in S] + \delta$.
\end{definition}

\begin{definition}[Bounded$(\epsilon, \delta)$-Differential Privacy)~\cite{dwork2006calibrating}] \label{def:dp_bounded}
A mechanism M is $(\epsilon, \delta)$ differentially private if every neighboring pair of datasets $D, D' \in U$ differing at most on the value of one data point in an attribute $A$, and all $S \subseteq Range (M)$, the following inequality holds:
$Pr[M(D.A) \in S] \leq e^\epsilon \cdot Pr[M(D'.A) \in S] + \delta$.

\end{definition}

The definitions guarantee that it is information-theoretically challenging to infer whether a data point belongs to either $D$ or $D'$ with certain confidence $\delta$. 
$\epsilon$ and $\delta$ are the privacy parameters. $\epsilon$ controls the privacy loss, and intuitively, $\delta$ is the probability that the privacy loss would be greater than $\epsilon$.
The smaller the parameters are, the stronger the privacy is. In practice, a typical value for $\epsilon$ is 1, and for $\delta$ is $\frac{1}{N}$ where N is the number of data points in the dataset. The notion ($\epsilon,\delta$)-DP turns to be $\epsilon$-DP when $\delta$ is 0.

A differentially private mechanism $M_f$ injects a noise to a query result $f$.
The sensitivity of $f$ controls how much difference in the output depends on the input.

\begin{definition} [Global Sensitivity] \label{def:gs2}
Let $f : D \rightarrow \mathbb{R}^d$.
\begin{itemize}
    \item Global sensitivity w.r.t. the presence of an item is $\Delta f= \max\limits_{D,D'} |f(D) - f(D')|$;

    \item  Global sensitivity w.r.t. the value of an item in an attribute $A$ is $\Delta^A f= \max\limits_{D,D'} |f(D.A) - f(D'.A)|$;

\end{itemize}
where $\max$ is computed over every neighboring pair of datasets $D,D'$ differing on the value of one item.

\end{definition}

Given the dataset $D$, and the query $f$, a mechanism $M_f$ returns a noisified result $f(D) + Y$, where $Y$ is the generated noise. The injected noise is randomly drawn from a particular distribution. One commonly used distribution for the numeric queries is the Laplace distribution $Lap(b, \mu)$. The probability density function of the Laplace distribution is $\frac{1}{2b} exp(-\frac{|x-\mu|}{b})$, where $\mu$ is the mean, and b is the scale factor. With $Lap(b, 0)$, we can obtain $\epsilon$-DP mechanism~\cite{dwork2014algorithmic}.

The classical design of DP mechanisms takes the parameters $\epsilon$ and $\delta$ as inputs and then carefully injects randomness to provide the privacy guarantee. This randomness can be achieved via sampling~\cite{Elkoumy21MineMe} or via changing the values of an attribute~\cite{mannhardt2019privacy,fahrenkrog2020pripel}.
Simultaneously, the DP mechanism tries to preserve the utility. However, this paradigm has been shifted recently since better privacy-utility tradeoffs can be achieved using a fine-grained analysis adapted to a specific mechanism~\cite{Wang19Subsampled,Abadi16DeepLearning,ballePrivacyAmplification}.

Recent studies~\cite{Wang19Subsampled,ballePrivacyAmplification} have shown that subsampling can be used in designing DP mechanisms that provide privacy amplification. 
Roughly speaking, if we apply an $(\epsilon, \delta)$ differentially private mechanism to a $\gamma$-proportion random subsample of the dataset, the output satisfies $((O(\gamma\epsilon),\gamma \delta))$-DP~\cite{ballePrivacyAmplification}. This paper adopts Poisson subsampling of the cases in an event log to provide more optimal use of the privacy budget~\cite{ZhuW19Poission}.

\subsection{Privacy-Preserving Process Mining}
\label{sec:pppm}
One of the earliest PPPM approaches was proposed by Manndart et al.~\cite{mannhardt2019privacy}, which provides differential privacy for process mining. They anonymize the query ``frequencies of directly-follows relations'' and the query ``the frequencies of trace variants''. Furthermore, PRIPEL~\cite{fahrenkrog2020pripel} adopts differential privacy to anonymize event logs. It embraces sequence enrichment for injecting noise to timestamps and other attributes of the event log. The above approaches accept three parameters: $\epsilon$, $k$, and $N$. $k$ is the cut-off frequency to filter out variants that occur less than $k$, and $N$ is the length of the longest trace in the anonymized logs. The output of the above approaches may contain variants that were never observed in the event log. Besides, some of the newly injected variants are impossible to occur for the anonymized process~\cite{fahrenkrog2021sacofa}.

SaCoFa~\cite{fahrenkrog2021sacofa} is another differentially private mechanism that achieves lower utility loss than the above approaches by using semantic constraints. This approach adopts DP to replace prefixes common in multiple cases with perturbed ones, given that the distance between the prefixes and the original one does not exceed a certain distance. However, this approach suffers from false negatives (traces that do not happen in the event log) and false positives (traces that occurred in the original log but do not appear in the anonymized log).

Elkoumy et al.~\cite{Elkoumy21MineMe} propose an approach that provides differentially private anonymization of event logs. They provide a privacy guarantee that the probability of singling out an individual in the anonymized log does not exceed a given threshold. They provide that guarantee  while preserving the same trace variant as the input log. They adopt oversampling as their noise injection mechanism. However, this approach results in high noise injection over the directly-follow frequencies between activities.

Other studies adopt group-based models, as in~\cite{rafiei2020tlkc,rafiei2021group,fahrenkrog2019pretsa,Batista21Uniformization}. These approaches suppress entire cases or individual events (activity instances). When a technique suppresses an activity instance in a case, it may introduce behavior (e.g., \ a directly-follows relation) that does not exist in reality. When this suppression occurs at scale, the anonymized log contains a high proportion of behavior that does not happen in the original log. In a patient treatment log used in the evaluation reported in this paper, existing k-anonymization techniques for process mining may lead to the suppression of 87\% of the activities~\cite{rafiei2020tlkc}.

Other studies of PPPM fall outside the scope of this paper, as they do not provide a mechanism to anonymize event logs. For example,
privacy-preserving process performance indicators have been proposed in~\cite{KabierskiFW21}. A risk quantification method has been presented in~\cite{VoigtFJKTMLW20}. Privacy-preserving continuous event data processing has been studied in~\cite{RafieiA21}. 
Other studies provided secure processing for distributed event logs in inter-organizational settings~\cite{ElkoumyFDLPW20Secure,ElkoumyFDLPW20Shareprom}.

\section{Approach}
\label{sec:approach}

\begin{figure*}[hbtp]
\vspace*{-2mm}
\centering
\includegraphics[width=1.6\columnwidth]{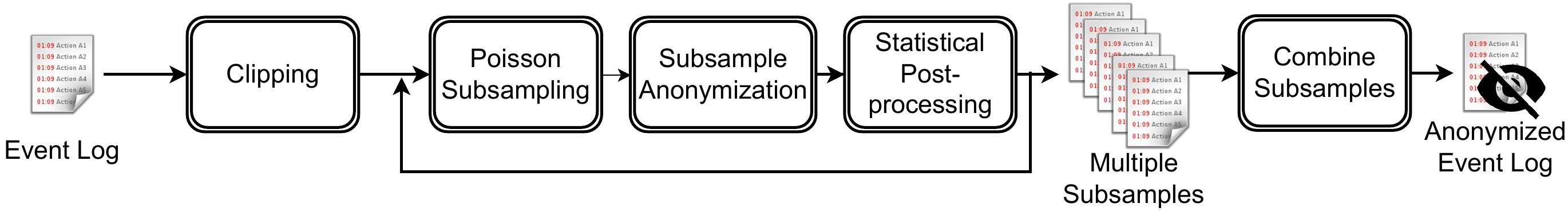}

\caption{ Approach }
	\label{fig:approach}
   \vspace*{-3mm}
\end{figure*}

We seek to anonymize an event log in such a way an attacker cannot single out an individual (e.g., a patient) in the anonymized event log. Accordingly, Libra applies differential privacy to the event log. To provide better usage of a given privacy budget $\epsilon$, Libra relies on privacy amplification by subsampling~\cite{ZhuW19Poission,ballePrivacyAmplification}.

More concretely, given an event log and a privacy budget $\epsilon$, Libra produces a differentially-private event log in 5 steps as outlined in Fig.\ref{fig:approach}. Some trace variants may be rare; keeping them can lead to singling out individuals. Accordingly, the first step clips trace variants below a clipping threshold C, estimated by the given privacy budget $\epsilon$. The second step constructs a Poisson subsample of the event log.  The third step is to anonymize subsamples.
Some of the anonymized cases may affect the utility. Therefore, Libra performs statistical post-processing of the anonymized subsamples to select the relevant traces. Libra repeats the above process to generate multiple subsamples while tracking the privacy budget by noise screening, and it composes the privacy budgets of the subsamples using Renyi Differential Privacy (RDP). Lastly, Libra combines the generated subsamples to construct the anonymized event log. In the following, we introduce the attack model and explain each step of Libra in turn. Furthermore, an algorithm is presented to formalize Libra's steps.

\subsection{Attack Model}
\label{subsec:attackModel}

We consider a setting where an event log publisher shares the log with an analyst who should not be able to link a case to an individual (e.g., a customer). An event log captures the activities of  process execution, as illustrated in Table~\ref{tbl:event_log}. Each log record is an event with an activity label, an identifier of the process instance (Case ID), and a timestamp. An event may have other attributes, such as resources. Libra anonymizes event logs from the control flow perspective, representing the traces of individuals under study, and from the time perspective. Thus, we assume the input event log has only the three attributes: Case ID, activity label, and timestamp. A trace represents the sequence of events  executed for a case.

\begin{definition} [Event Log, Event, Trace]\label{def:event_log}
	An event log $L= \{e_1, e_2, ..., e_n\}$ is a record of a set of events $e=(i,a,ts)$, each captures an activity execution $a$ for a process instance (case) i, at timestamp $ts$. A trace $t=\langle e_1, e_2, ..., e_m\rangle$ captures the sequence of events in L of a case i. An event log can be represented as a set of traces $L= \{t_1, t_2, ..., t_z\}$.
\end{definition}

We wish to protect the privacy of individuals (customers in this setting) from an analyst (the attacker in this setting) who seeks to determine if an individual participated in the log or not. Based on that, the attacker can link a case to an individual.
An attacker can determine the existence of an individual in the released event log by using either the sequence of activities of a trace (control flow) or the execution time of a particular activity.
Specifically, we seek to anonymize log L in such a way to prevent the attacker from achieving the following goals:

\begin{itemize}
    \item $h_1$: Distinguishing whether an individual has participated in the log or not through their execution control flow. 
    
    \item $h_2$: Determining the execution timestamp of an activity.
\end{itemize}

    


We assume that the set of activity labels is public 
and that the execution of traces is independent. Following, we introduce the concept of a differentially-private event log that prevents an attacker from achieving the goals $h_1$ and $h_2$.

\begin{definition} [Differentially-Private Event Log]\label{def:DPEL}
    Given an event log L as defined in Def.~\ref{def:event_log}, an anonymized log L$'$ is said to be $\epsilon$-differentially private if: (1) it provides $\epsilon$-differential privacy over the control flow of traces; (2) the timestamp attribute is $\epsilon$-differentially private.
\end{definition}

\subsection{Clipping Rare Traces}
As explained above, an individual could be singled out through their control flow if they have a unique sequence of activities or trace. The set of the unique traces in the event log is called trace variants.

\begin{definition} [Trace Variants] \label{def:case_variants}
A trace variant of an event log $L$ is a sequence of activities $\langle\, a_1, a_2, ..., a_k \,\rangle$ such that there is a trace $\langle\, e_1, e_2, ..., e_k \,\rangle$ of $L$ such that $\forall j \in [1..k]: \; id(e_j) = a_j$, where $id(e)$ is the case ID of the event $e$.
\end{definition}

It is possible to single out individuals using their trace variants, even if the variant is not unique, but exists a few times in the log. We call such a trace variant a \textit{Rare trace variant}~\cite{ChaudhuriM06RareValues}. Observing a rare trace variant is problematic because it is executed for a group of few individuals. Observing such a trace may increase the attacker's confidence about this group of individuals. Chaudhuri et al.~\cite{ChaudhuriM06RareValues} define rare values (rare trace variants) in a data sample w.r.t ($\epsilon,\delta$)-DP guarantee.

\begin{lemma}[Rare Trace Variants~\cite{ChaudhuriM06RareValues}] \label{lemma:rare_values}
A sample that provides ($\epsilon,\delta$)-differential privacy guarantee should not contain rare trace variants that happen less than $1/\epsilon  \; log(\frac{2k}{\delta})$, where $k$ is the number of trace variants in the log.
\end{lemma}

Since Libra provides privacy amplification via subsampling, any log sample that provides ($\epsilon,\delta$)-DP guarantee should not contain rare trace variants~\cite{ChaudhuriM06RareValues}.
If a rare trace variant exists in the released log, it may lead to a privacy breach of $O(\frac{i \epsilon}{log(2k/\delta)})$, where $i$ is the number of rare traces in the log~\cite{ChaudhuriM06RareValues}.
Libra clips trace variants that happen less than $C=1/\epsilon \; log(\frac{2k}{\delta})$.
It does so by filtering out all the trace instances of the trace variants that happen $< C$.

Libra does not apply when the log is a collection of unique traces (i.e., each case is of a unique trace variant). The reason is that if every trace in the log is unique, then a sample of any size, even one, violates the privacy guarantees.


\subsection{Event Log Subsampling}
Our goal is to prevent an attacker from determining the existence of an individual in the anonymized log. 
To this end, we bring privacy amplification by subsampling~\cite{ballePrivacyAmplification,ZhuW19Poission} into action.

\begin{definition} [Subsampling]\label{def:subsample}
    Given an event log L of z traces, a subsampling procedure selects a random set of traces from the uniform distribution of traces of L of size m. The ratio $\gamma := \frac{m}{z}$ is the sampling ratio of the subsampling procedure.
\end{definition}

Given Def.~\ref{def:dp}, DP works by adding or removing an item. Poisson sampling works naturally with that property~\cite{ZhuW19Poission}.

\begin{definition}[Poisson Sample~\cite{ZhuW19Poission}]\label{def:poisson_smple}
Given an event log L, a PoissonSample procedure returns a subset of traces of the event log $\{ t_i| \sigma_i = 1, i \in [1..z] \}$ by sampling $\sigma_i \sim Ber(\gamma)$ independently for $i=1,2,...,z$.
    
\end{definition}

The above procedure is equivalent to performing subsampling without replacement with sampling rate$ \sim Binomial(\gamma,z)$.
A binomial distribution converges to a Poisson distribution with parameter $\lambda$ at the limit of $k \rightarrow \infty, \gamma \rightarrow 0 $ while $\gamma z \rightarrow \lambda$.

Libra applies Poisson Subsampling to the input event log and generates multiple subsamples. 
 Each subsample contains entire traces rather than subtraces. That prevents the subsampling step from introducing trace variants that do not exist in the log. The next step explains the anonymization of each subsample.

\subsection{Subsamples Anonymization} \label{sec:subsamp_anony}

At this step, Libra provides an $\epsilon$-DP guarantee for each subsample. 
Privacy amplification is achieved by applying a DP mechanism to a subsample of the dataset~\cite{ballePrivacyAmplification,Wang19Subsampled}.


\begin{lemma} [Privacy Amplification~\cite{ballePrivacyAmplification,ZhuW19Poission}]
If a mechanism M is ($\epsilon, \delta$)-DP, then a subsampled mechanism M $\circ$ PoissonSubsample provides ($\epsilon', \delta'$)-DP with $\epsilon'= log(1+\gamma(e^\epsilon -1 ))$ and $\delta' = \gamma \delta$.

\end{lemma}

Following, we explain how Libra applies differential privacy for each subsample (M $\circ$ PoissonSubsample). The goal of subsamples anonymization is to prevent attacks $h_1$ and $h_2$.

\begin{definition}[Differentially-Private Subsample]\label{def:DP_subsamples}
Given a PoissonSample $S$ as defined in Def.~\ref{def:poisson_smple}, the output of a mechanism $S'=M(S)$ is said to be $\epsilon$-differentially private if: (1) it provides $\epsilon$-differential privacy over the set of trace variants; (2) the timestamp is $\epsilon$-differentially private.
\end{definition}

To anonymize the subsamples, we could use any of the DP mechanisms in the literature~\cite{mannhardt2019privacy,fahrenkrog2020pripel,fahrenkrog2021sacofa,Elkoumy21MineMe}. We use the mechanism presented in~\cite{Elkoumy21MineMe} because it does not introduce trace variants that do not exist in the original log. It employs a guessing advantage to estimate the differential privacy parameter $\epsilon$. Then, it injects the noise over a log representation called DAFSA. The noise injection for trace variants is done by oversampling traces. Libra adopt the approach in~\cite{Elkoumy21MineMe} to sample traces (adding or deleting traces) instead of oversampling. Also, we provide it with an $\epsilon$ value rather than a guessing advantage threshold. In the evaluation section, we measure the impact of privacy amplification on utility loss, and we compare  both Libra and the one in~\cite{Elkoumy21MineMe}.

To protect against linking a case to its individual via the cycle time of an activity ($h_2$), we inject noise drawn from a Laplace distribution quantified by a given privacy budget $\epsilon$. Therefore, we anonymize two components of the timestamps: (1) case start time, which is the timestamp of the first event of the case; (2) the execution timestamp of every other event after that. In order to inject the noise to the start time, we introduce the relative start time of the case, which is the difference between the case start time and the first start time in the log. We inject random noise quantified by $\epsilon$ to the relative start time. Both the relative start time and the generated noise are measured in days.

On the other hand, we aim to protect the cycle time of each event, which is the difference between its execution timestamp and the execution timestamp of its successor event. We inject randomly generated noise quantified by $\epsilon$ to the cycle time. Both the cycle time and the generated noise are measured in minutes. 
After injecting noise as mentioned above, we transform the anonymized relative start time and cycle time to timestamps again via addition. At the end of this step, we have a differentially private subsample of the log (M $\circ$ PoissonSample).

\subsection{Statistical Post-processing of SubSamples}\label{sec:post-process}
Anonymization perturbs the utility of event logs. Libra selects statistically significant traces out of the anonymized log to provide higher utility of the differentially private event log. This selection process is a post-processing step to the anonymized subsamples. The result of a post-processing step of differentially-private subsamples of a log provides the same differential privacy guarantees~\cite{dwork2014algorithmic}.
\begin{proposition}[Differential Privacy Under Post-processing~\cite{dwork2014algorithmic}]\label{prop:postprocessing}
A post-processing procedure P of a subsampled log L$'$ provides ($\epsilon,\delta$)-DP guarantees, if and only if L$'$ is ($\epsilon,\delta$)-differentially private.
\end{proposition}

The proof of Prop.~\ref{prop:postprocessing} is in~\cite{dwork2014algorithmic} (c.f. Proposition 2.1).

Given the differentially private  Poisson subsample of the log, the post-processing of the selected subsample provides ($\epsilon,\delta$)-DP guarantees. 
Libra uses statistical post-processing of the subsamples to pick the most relevant traces and reduce the utility loss.
These  relevant traces are assessed by a trace abstraction function $\psi : 2^\omega \rightarrow \chi$, where $\chi$ is the domain of information extracted from a trace~\cite{MartinSampling}. This information can be related to the sequence of activities in a trace and the frequencies of activities.
Bauer et al.~\cite{MartinSampling} provide a log sampling mechanism that adopts a series of binomial experiments and picks traces that provide new information while being discovery sufficient with probability $\rho$. Libra adopts the work in~\cite{MartinSampling} to pick the relevant traces out of the differentially private subsamples. A trace $\tau$ provides new information if its abstraction is far from the union of the abstraction, jointly derived from the subsamples. That should happen within a distance $\omega$ of the union abstraction. Thus, we consider the predicate~\cite{MartinSampling}:

\begin{equation}\label{eqn:predicate}
\footnotesize
\pi^\omega(S'_e,\tau) \leftrightarrow d\bigr(\psi(\tau), \bigcup_{\tau' \in S'_e} \psi(\tau')\bigr)> \omega\enspace,
\end{equation}

where $\pi$ is the picked sample, $S'_e$ is the differentially private subsample, and $\psi$ is the used abstraction. A subsample is discovery sufficient w.r.t. an abstraction $\psi$, a distance parameter $\omega$, and probability $\rho$ as follows:

\begin{definition}[Discovery Sufficiency~\cite{MartinSampling}]
A DP-subsample $S'_e \subset S'$ is ($\rho, \omega, \psi$)-discovery sufficient, if there is a newly picked trace $\tau: \tau \in (S'\setminus S'_e)$ that:
$p_\pi (S'_e, \tau) = P(\pi (S'_e,\tau)=1 < \omega$,
where p is a probability measure.
\end{definition}

At the end of this step, Libra has ($\epsilon,\delta$)-differentially private subsamples of the log that have statistically relevant traces.

\subsection{Combining Subsamples}
 \vspace*{-1mm}
In order to construct an ($\epsilon,\delta$)-differentially private event log with the number of traces as close as possible to the number of traces in the original log, $z$, Libra repeats the differentially private subsampling operation for a number of times equals $\eta=\gamma z$.
The repetitive access of the log via DP-subsampling is identical to the composition of the multiple differentially-private mechanisms (one mechanism for each DP subsample).

\begin{theorem}[Differential Privacy Composition~\cite{dwork2014algorithmic}] \label{theorem:composition}
Let $M_1$ and $M_2$ be $\epsilon_1, \epsilon_2$-differentially-private mechanisms. Then, the combination of the mechanisms $M_{1,2}(L) = (M_1(L),M_2(L))$ is $\epsilon_1+\epsilon_2$-differentially private.
\end{theorem}
The proof of Theorem~\ref{theorem:composition} is in~\cite{dwork2014algorithmic} (c.f. Theorem 3.14).

Indeed, the privacy parameters $\epsilon,\delta$ degrade due to the composition. This degradation is compensated by the privacy amplification. One way to estimate $\epsilon,\delta$ values after composition is using Renyi Differential Privacy (RDP)~\cite{Mironov17RDP}.


\begin{definition}[Renyi Differential Privacy~\cite{Mironov17RDP}]
A mechanism M is said to be ($\alpha,\epsilon$)-RDP with order $\alpha\in(1,\infty)$ if for every neighboring datasets D,D$'$:
\begin{equation*}
\footnotesize
D_\alpha (M(D) || M(D')) =  \frac{1}{\alpha-1} log \biggr( E_{\theta ~ M(D')} [ ( \frac{P_{M(D)} (\theta)}{P_{M(D')}  (\theta)} )^\alpha ] \biggr) \leq \epsilon. 
\end{equation*}
\end{definition}



$\alpha$ is the Renyi divergence order between the two datasets $D, D'$. Mironov~\cite{Mironov17RDP} provides an estimation of the equivalent $\epsilon$ of the RDP as a function of $\alpha$.
The estimated $\epsilon$ for the Laplace distribution equals:
\begin{equation} \label{eq:eps_alpha}
\footnotesize
    \epsilon_{Laplace}  (\alpha)= \frac{1}{\alpha-1}\; log \biggr( \frac{\alpha}{2\alpha -1} e^{\frac{\alpha-1}{b}} + \frac{\alpha-1}{2\alpha-1} e^{\frac{-\alpha}{b}}  \biggr)
\end{equation}


The $\epsilon$ estimated in Eq.~\ref{eq:eps_alpha} can be passed to the Laplace mechanism to draw the noise and anonymize the subsamples. 
Zhu et al.~\cite{ZhuW19Poission} estimate the amplified $\epsilon'$ after Poisson subsampling and composition using RDP as:



\begin{eqnarray}\label{eq:eps_subsample}
\footnotesize 
\epsilon'_{M \circ PoissonSubsample}(\alpha) \leq \frac{1}{\alpha-1} log \biggr((1-\gamma)^{\alpha-1} ( \alpha \gamma - \alpha+1) \nonumber\\
\binom{\alpha}{2} \gamma^2 (1-\gamma)^{\alpha-2} e^{\epsilon(2)} \nonumber\\
+ 3 \mathlarger{\sum}_{j=3}^{\alpha} \binom{\alpha}{j}  (1-\gamma)^{\alpha-j} \gamma^j   e^{(j-1) \epsilon (j)}  \biggr)
\end{eqnarray}

\subsection{Libra Anonymization Algorithm}
 \vspace*{-1mm}
We exploit the above observation as formalized in Alg.~\ref{alg:approach}. Libra takes as input an event log L, an order of RDP $\alpha$, a differential privacy parameter $\delta$, and a sampling ratio $\gamma$. The algorithm estimates the $\epsilon$ used to draw noise from the Laplace distribution using Eq.~\ref{eq:eps_alpha}. Then, it sets $k$ to equal the count of trace variants in $L$. The algorithm estimates the clipping threshold $C$ as explained in Lemma~\ref{lemma:rare_values}.  The algorithm uses the estimated C to filter out trace variants that occur less than $C$ from the log $L$. We set $z$ to equal the number of cases in the filtered log $\hat{L}$. Then, the algorithm estimates $\eta$, the number of subsamples needed to construct an event log with as many cases as the filtered log $\hat{L}$. After that, the algorithm performs Poisson sampling over $\hat{L}$. Given the Poisson subsample $S$, the algorithm generates an anonymized sample $S'$ as explained in Sec.~\ref{sec:subsamp_anony}. Then, it picks the statistically relevant traces from the anonymized sample $S'$ as explained in Sec.~\ref{sec:post-process}. The algorithm generates $\eta$ subsamples and combines them to generate $L'$. Also, the algorithm reports the amplified $\epsilon'$ after the composition of privacy budgets of subsamples using Eq.~\ref{eq:eps_subsample}.

\vspace*{-2mm}
\begin{algorithm}

\scriptsize 
\hspace*{\algorithmicindent} \textbf{Input:}~$L$:~Event~Log, \\
                            $\alpha$: order of Renyi Differential Privacy,\\
$\delta$: Differential Privacy parameter,\\
$\gamma$: Sampling Ratio.\\
 \hspace*{\algorithmicindent} \textbf{Output:}~$L'$: $\epsilon'$-Differentially~Private~Event~Log,\\
 $\epsilon'$: the amplified differential privacy guarantee. \\

$\epsilon \leftarrow$ EstimateEpsilon($\alpha$); \Comment{Estimate $\epsilon_{Laplace}$ using Eq.~\ref{eq:eps_alpha}.}\\ 

$k \leftarrow $ CountVariants($L$); \Comment{Set $k$ to the count of trace variants in $L$.}\\

$C\leftarrow (1/\epsilon)\; log(2k/\delta)$;\Comment{Estimate clipping threshold based on Lemma~\ref{lemma:rare_values}.} \\

$\hat{L} \leftarrow$ Filter(L, C); \Comment{Filter out trace variants with frequency below C.}\\
$z \leftarrow $ CountCases($\hat{L}$); \Comment{Set $z$ to the count of cases in $\hat{L}$.}\\

$\eta \leftarrow \gamma z$; \Comment{Calculate the count of subsamples.}\\

$L' \leftarrow 0$;\\
 \While{ i $< \eta$ } {
S= PoissonSample($\hat{L}, \gamma$); \Comment{Perform Poisson subsample as defined in Def.~\ref{def:poisson_smple}.}\\

$S' \leftarrow$ Anonymize($S,\epsilon$); \Comment{Anonymize the subsample as explained in Sec.~\ref{sec:subsamp_anony}.}\\

$S'_e \leftarrow $; StatisticalPost-processing($S'$); \Comment{Perform  statistical post-processing to select relevant cases as explained in Sec.~\ref{sec:post-process}.}\\

$L' \leftarrow L' \cup S'_e$;

i++ \;
  
 }

 $\epsilon' \leftarrow$ EstimateComposition($\alpha, \epsilon,\gamma$); \Comment{Estimate $\epsilon'$ using Eq.~\ref{eq:eps_subsample}.} \\
 
\Return{ $L',\epsilon'$}
 
 \caption{The Libra Anonymization Algorithm}
 \label{alg:approach}

\end{algorithm}
\vspace*{-2mm}

Alg.~\ref{alg:approach} employs a differentially private mechanism to anonymize every Poisson subsample and performs post-processing of subsamples to pick the relevant traces. Since the post-processing depends only on the anonymized samples, the output of the post-processing (DP-subsamples) is differentially private, c.f., Prop.~\ref{prop:postprocessing}. After that, the algorithm uses differentially private composition to combine the differentially-private subsamples to construct the anonymized event log. Thus, the algorithm's output is differentially private.

\section{Evaluation}
\label{sec:experiment}
To address the research problem stated in Sect.~\ref{sec:intro}, Libra injects noise to provide ($\epsilon,\delta$)-DP guarantees in two ways: (1) by filtering out rare trace variants and sampling traces in the log; (2) by introducing time shifts to the event timestamps. Filtering and noise injection affect the utility of the anonymized logs. To measure this utility loss, we compare the anonymized event logs against the original logs using a distance measure. Also, we compare the performance (execution time) of Libra against state-of-the-art baselines.

The evaluation reported below is driven by the following research questions:
\begin{enumerate}[label=\textbf{RQ\arabic*.}]
    \item \label{res:rq:baseline} Does Libra outperform the state-of-the-art baselines in terms of the output utility?
    \item \label{res:rq:performance} What is the difference between Libra and the state-of-the-art in terms of computational efficiency?
\end{enumerate}

\subsection{Evaluation Measures}

Given an event log, mainstream process mining tools generally produce a Directly-Follows Graph (DFG) to capture the behavior of the process for the purpose of process discovery and analysis. In addition to being used as a visualization technique on its own, DFGs are the starting point for several algorithms to discover process models from event logs. Accordingly, and in line with prior work, we measure  utility loss over the DFG of an event log. A DFG can be seen as a function that maps each pair of consecutive activities in the log (i.e.\ each directly-follows relation) to its frequency. This function can be seen as a frequency distribution. Accordingly, to measure utility loss over DFGs, we use the \textit{Earth Movers' Distance}(EMD)~\cite{ramdas2017wasserstein}. The EMD distance between two distributions $u$ and $v$ is the minimum cost of transforming $u$ into $v$. The cost is the weight of the distribution to be moved, multiplied by the distance it moves. Formally:

\begin{equation}\label{eq:emd}
\footnotesize
EMD (u,v)= \inf_{\pi \in \Gamma(u,v)} \int_{\mathbb{R}\times \mathbb{R}} |x - y | d\pi(x,y),
\end{equation}
 where $\Gamma(u,v)$ is the set of distributions on $\mathbb{R}\times \mathbb{R}$ whose marginals are $u$ and $v$. 
 
In order to measure the computational efficiency of Libra, we measure the wall-to-wall run time experiment. We measure the time between submitting an event log to the approach and receiving its anonymized version.

\subsection{Datasets}
In our evaluation, we rely on the publicly available event logs of  4TU Centre for Research Data\footnote{\url{https://data.4tu.nl/}} as of February 2022. The selected logs and their characteristics are presented in the supplementary material\footnote{\url{https://doi.org/10.5281/zenodo.6376761}}. 


		
	
		
		
		


\subsection{Experiment Setup}
We implement the proposed model as part of Libra\footnote{\url{https://github.com/Elkoumy/Libra}} prototype. We run the experiment on a single machine with  AMD Opteron(TM) Proc 6276 and 32 GB RAM. We time out any experiment at 24 hours.
Also, in our experiment, we consider only the end timestamp to calculate the relative time of an event for simplicity. The approach still holds DP guarantees for logs with start and end timestamps.
We fix the parameters b=2, $\gamma$=0.05, and $\delta=10^{-4}$, and we evaluate the approach for different values of $\alpha$. For an empirical evaluation of the relation between $\gamma$, $\alpha$, and $\epsilon$ under Poisson subsampling, we refer to~\cite{ZhuW19Poission}.

We compare our approach against the state-of-the-art. The studies that address the same problem are~\cite{rafiei2020tlkc,rafiei2021group,mannhardt2019privacy,fahrenkrog2020pripel,fahrenkrog2021sacofa,Elkoumy21MineMe}. We exclude the work in~\cite{rafiei2020tlkc,rafiei2021group} from our experiment because the interpretation of the k-anonymity parameters and DP are different. The studies in~\cite{mannhardt2019privacy,fahrenkrog2020pripel,fahrenkrog2021sacofa,Elkoumy21MineMe} provide DP guarantees. The work in~\cite{mannhardt2019privacy} anonymizes two types of queries, but the output is not an event log. PRIPEL~\cite{fahrenkrog2020pripel} adopts~\cite{mannhardt2019privacy} for trace variant anonymization. We compare Libra against~\cite{fahrenkrog2020pripel}. SaCoFa~\cite{fahrenkrog2021sacofa} provides trace variant anonymization but does not consider timestamp anonymization. Therefore, we consider~\cite{fahrenkrog2021sacofa} only in EMD of frequency annotated DFG experiments. Both PRIPEL and SaCoFa take three input parameters namely, $\epsilon$, k, and N. For the pruning parameter k, we run several experiments with different values of k(0.5\%, 1\%,5\% of the cases), and we select the best result. For the maximum trace length N, we set N to the average trace length of the log. Amun~\cite{Elkoumy21MineMe} anonymizes both the trace variants and the timestamps. It accepts the guessing advantage probability $\delta$, rather than $\epsilon$. Thus, we set $\delta$ to the values (0.01, 0.075, 0.4) that give an average $\epsilon$ close to the $\epsilon$ considered in the experiments. We use the EMD to compare the anonymized log against the original one for the selected approaches.

\subsection{Results}
\begin{table*}[hbtp]

	\caption{Earth Movers' Distance for the output of different anonymization approaches. A ``-'' means that the approach ran out of memory or timed out, and A ``N/A'' means the engine returns an empty log.}
	\label{tbl:emd}
 \scriptsize

\centering
	\begin{tabular}{ p{1.3cm} p{0.8cm} p{0.8cm} c c c cc c c   }
		\toprule
        \multirow{2}{*}{Log}		&	\multirow{2}{*}{$\alpha$} 	&	\multirow{2}{*}{$\epsilon'$}	&	 \multicolumn{4}{c}{EMD Freq} &	 \multicolumn{3}{c}{EMD Time}	\\
        	\cmidrule(lr){4-7} \cmidrule(l){8-10}	&		&		&	   $Amun_o$	&	  $Libra$	&	  $PRIPEL$	&	  $SaCoFa$	&	  $Amun_o$	&	  $Libra$	&	 $PRIPEL$		\\
        \hline

\multirow{3}{*}{BPIC12}	&	2	&	0.04	&	-	&	1036	&	-	&		\textbf{1007}	&	-	&	\textbf{19315}	&	-	\\
	&	10	&	0.37	&	38442	&		\textbf{997}	&	-	&	1007	&	15068248	&	\textbf{19228}	&	-	\\
	&	100	&	4.7	&	1293	&		\textbf{1005}	&	1230	&	1007	&	54328	&	\textbf{19224}	&	\textbf{19224}	\\\hline
																			
\multirow{3}{*}{BPIC13}	&	2	&	0.04	&	18853	&		\textbf{3847}	&	4954	&	4947	&	9520973	&	\textbf{145272}	&	197564	\\
	&	10	&	0.37	&	19573	&		\textbf{3586}	&	4952	&	4948	&	10552209	&	\textbf{132827}	&	197564	\\
	&	100	&	4.7	&	4436	&		\textbf{3492}	&	4952	&	4952	&	1208926	&	\textbf{127752}	&	197564	\\\hline
																			
\multirow{3}{*}{BPIC14}	&	2	&	0.04	&	-	&		\textbf{494}	&	-	&	544	&	-	&	\textbf{7088}	&	-	\\
	&	10	&	0.37	&	-	&		\textbf{466}	&	-	&	544	&-	&	\textbf{6970}	&-		\\
	&	100	&	4.7	&	1013	&		\textbf{464}	&	-	&	507	&	128218	&	\textbf{6944}	&	-	\\\hline
																			
\multirow{3}{*}{BPIC17}	&	2	&	0.04	&	-	&		\textbf{1563}	&	-	&	2785	&	-	&	\textbf{48254}	&-		\\
	&	10	&	0.37	&	-	&		\textbf{1349}	&	-	&	2789	&	-	&	\textbf{41190}	&-		\\
	&	100	&	4.7	&	\textbf{939}	&		1333	&	-	&	2790	&	148867	&	\textbf{40864}	&-		\\\hline
																			
\multirow{3}{*}{BPIC20}	&	2	&	0.04	&	-	&		\textbf{91}	&	128	&	125	&	-	&	\textbf{15602}	&	24422	\\
	&	10	&	0.37	&	342	&		\textbf{82}	&	128	&	127	&	629659	&	\textbf{13920}	&	24422	\\
	&	100	&	4.7	&		\textbf{79}	&		\textbf{79}	&	128	&	127	&	58201	&	\textbf{13504}	&	24430	\\\hline
																			
\multirow{3}{*}{Hospital}	&	2	&	0.04	&	-	&	N/A	&	-	&	\textbf{35}	&	-	&	N/A	&-		\\
	&	10	&	0.37	&	-	&	\textbf{35}	&	-	&	\textbf{35}	&	-	&	\textbf{2507}	&	-	\\
	&	100	&	4.7	&	-	&	\textbf{35}	&	\textbf{35}	&	\textbf{35}	&	-	&	2507	&	\textbf{2469}	\\\hline
																			
\multirow{3}{*}{Sepsis}	&	2	&	0.04	&	 1120	&	N/A	&	\textbf{75}	&	117	&	214068	&	N/A	&	\textbf{4644}	\\
	&	10	&	0.37	&	1061	&	123	&	\textbf{120}	&	121	&	205646	&	\textbf{6218}	&	\textbf{6218}	\\
	&	100	&	4.7	&	234	&	123	&	\textbf{120}	&	\textbf{120}	&	21661	&	\textbf{6218}	&	\textbf{6218}	\\\hline
																			
\multirow{3}{*}{Unrine.}	&	2	&	0.04	&	1021	&	120	&	298	&	\textbf{42}	&	2921865	&	\textbf{58576}	&	68245	\\
	&	10	&	0.37	&	169	&	95	&	292	&	\textbf{56}	&	547254	&	\textbf{49005}	&	68234	\\
	&	100	&	4.7	&	31	&	81	&	292	&	\textbf{64}	&	52284	&	\textbf{37381}	&	68234	\\

\bottomrule

	\end{tabular}
	\vspace*{-3mm}
\end{table*}

\begin{table}[hbtp]
	
	\caption{Execution time experiment. The time is measured in seconds for an  $\epsilon'=0.37$. A ``-'' means that the approach ran out of memory or timed out.}
	
	\label{tbl:time}
\scriptsize
\centering
	\begin{tabular}{c  c c c c}
		\toprule
dataset	&		$Libra$ &	$Amun_o$	&	$PRIPEL$&	\\
\hline
BPIC12	&	90	&	503	&	-	\\
BPIC13	&	49	&	115	&	306	\\
BPIC14	&	323	&	-	&	-	\\
BPIC17	&	212	&	-	&	-	\\
BPIC20	&	70	&	170	&	330	\\
Hospital	&	90	&	-	&	-	\\
Sepsis	&	10	&	80	&	24	\\
Unrine.	&	9	&	6.4	&	2	\\

\bottomrule

	\end{tabular}
	
\end{table}

We measure the EMD between the DFG of the anonymized log and the DFG  of the original log. We report the differences in terms of the frequency and time (measured in hours).
Table~\ref{tbl:emd} shows the experimental results using the EMD distance. 
$\alpha$ refers to the RDP parameter, and $\epsilon'$ refers to the equivalent DP parameter after the amplification and composition. The best result for every input $\alpha$ is in bold. Libra outperforms the state-of-the-art baselines in most of the logs, because the privacy amplification reduces the amount of injected noise, and hence achieves a lower utility loss.

Conversely, Libra has a lower frequency EMD than $Amun_o$ (\cite{Elkoumy21MineMe}) because the latter always injects positive noise, which affects the utility. Libra outperforms both PRIPEL and SaCoFa over the frequency EMD because for a given $\epsilon'$-DP guarantee, Libra needs an $\epsilon > \epsilon'$ due to privacy amplification. On the other hand, in event logs with many rare cases (Lemma~\ref{lemma:rare_values}) such as Sepsis and Urineweginfectie logs. Libra does not outperform the state-of-the-art because it adopts clipping to get rid of rare trace variants. Due to clipping, Libra sometimes return an empty log, e.g., for Sepsis log with $\alpha=2$ and Hospital log with $\alpha=2$. Also, Libra has a lower utility loss over the anonymized timestamps (measured by EMD time), which happens due to privacy amplification.

We evaluate the processing efficiency of Libra via a wall-to-wall run time experiment. Table~\ref{tbl:time} presents the experiments for $\alpha=10$ ($\epsilon'=0.37$). The time is measured in seconds. We exclude SaCoFa from this experiment as it provides only trace variant anonymization. The run time increases with the increase of the log size. The above experiments have been performed using a single thread to avoid adding other variables to the experiments.

We acknowledge that Libra is not suitable for event logs with many rare trace variants (unstructured event logs). In such a case the approach filters out most of the cases, and sometimes it returns an empty event log. Also, the above observations are based on a limited population (8 event logs). However, we selected the logs from a broader real-life event log population.


\section{Conclusion and Future Work}
\label{sec:conclusion}

This paper proposed a differentially private mechanism to anonymize event logs for process mining. 
While previous proposals in this field rely purely on noise injection, the approach proposed in this paper additionally employs subsampling to achieve stronger privacy guarantees with the same level of utility loss, or conversely, less utility loss for the same privacy guarantee (privacy amplification).
The empirical evaluation shows that the privacy amplification effect leads to significant reductions of utility loss, particularly when it comes to anonymizing the frequency of distribution of case variants in a log (i.e.\ control flow anonymization) and to a lesser extent when it comes to anonymization of event timestamps.


A limitation of Libra is that it is not suitable for event logs with a high proportion of infrequent trace variants. In such use cases, Libra simply filters out most of the traces and may lead to empty outputs. A possible future research avenue is to address this limitation by applying summarization techniques to replace groups of similar infrequent trace variants with a single prototypical trace variant, which would then have a frequency equal to the sum of the frequencies of the trace variants in the group.
A second limitation of Libra is that it only anonymizes three columns of an event log (case ID, activity label, and timestamp). Real-life log may contain additional event attributes, e.g. resources, cost, location. Another avenue for future work is to extend the privacy composition mechanism to include the anonymization of additional attributes.

\bibliographystyle{splncs04}
\bibliography{PrivacyAmplification}

\end{document}